\documentclass[twocolumn]{IEEEtran}
\usepackage{amssymb}
\usepackage{graphicx,cite,epsfig,amssymb,amsmath,epstopdf,algorithm,algorithmic,subfigure}
\usepackage{epstopdf}

\begin{document}

\title{Spectrum Intelligent Radio: Technology, Development, and Future Trends}

\author{\IEEEauthorblockN{Peng~Cheng, Zhuo~Chen, Ming~Ding, Yonghui~Li,
Branka~Vucetic, and Dusit~Niyato}}

%\thanks{P.~Cheng, Y.~Li, and B.~Vucetic are with the School of Electrical and Information Engineering, the University of Sydney, Australia, (e-mail: peng.cheng@sydney.edu.au; yonghui.li@sydney.edu.au; branka.vucetic@sydney.edu.au). Z.~Chen and M.~ Ding are with CSIRO DATA61, Australia (e-mail:zhuo.chen@ieee.org; ming.ding@data61.csiro.au).}}

\markboth{Accepted by IEEE Communications Magazine}{}

\maketitle

\begin{abstract}
The advent of Industry 4.0 with massive connectivity places significant strains on the current spectrum resources, 
and challenges the industry and regulators to respond promptly with new disruptive spectrum management strategies. 
The current radio development, 
with certain elements of intelligence, 
is nowhere near showing an agile response to the complex radio environments. 
Following the line of intelligence,
we propose to classify spectrum intelligent radio into three streams: 
classical signal processing, 
machine learning (ML), 
and contextual adaptation. 
We focus on the ML approach, 
and propose a new intelligent radio architecture with three hierarchical forms: 
perception, understanding, and reasoning.  The proposed perception method achieves fully blind multi-level spectrum sensing. The understanding method accurately predicts the primary users' coverage across a large area, and the reasoning method performs a near-optimal idle channel selection.     
  Opportunities, challenges, and future visions are also discussed for the realization of a fully intelligent radio.
\end{abstract}
%On  this  basis,  we  propose  a  new  architecture  formachine  learning-based  technology  with  three  hierarchica

%Machine learning-based intelligent cognitive radio (CR) is positioned to unlock the full spectrum potential, and pave the way for the implementation of future wireless networks with smart spectrum sharing.

\section{Introduction}
Industry 4.0~\cite{Industry4}, 
also known as the fourth Industrial Revolution, 
integrates device-enabled information collection and decision-making in the process of industrial production. 
%It aims to achieve a full automation of machines, 
%significantly boosting the productivity of human society. 
This paradigm-shifting disruption sets its cornerstone in massive wireless device connectivity, 
where densely populated sensors can observe, analyze, 
and communicate with the physical world to optimize operations and decisions. 
The massive machine-type communication (mMTC) in the emerging fifth-generation (5G) era promises to support one million devices per $\mbox{km}^2$ in the future. 
However, 
with the full swing of Industry 4.0 reaching everywhere, 
the anticipated device density could even surpass this target, 
particularly in large-scale smart factories.

A fundamental question naturally arises: 
\emph{How to manage this massive wireless access under the constraint of limited spectrum resources?}
%A quick reality check tells us that the sub-3\,GHz spectrum suitable for battery-powered low-cost devices has been crammed with many existing applications. 
%such as the legacy GSM networks working at 1$\sim$2\,GHz and the legacy Wi-Fi networks working at 2.4\,GHz. 
The traditional static radio spectrum management methods, 
based on interference-free frequency pre-allocation, 
cannot reflect the dynamics of real-time supply-and-demand, 
and thus creating the issue of spectrum scarcity. 
To support a large volume of devices in limited frequency spectrum, 
dynamic spectrum management (DSM)~\cite{CR} has been widely regarded as the most efficient solution.
%enabling more concurrent radio communications with given spectrum resource. 
In this scenario, 
secondary users (SUs)
%\footnote{In the future mMTC, the concepts of primary and secondary users and licensed spectrum could dwindle. Instead, all the devices, or a subset of the devices, could be treated equally in terms of spectrum access priority. As a result, decentralized and self-organized CR is more relevant.}
monitor the spectrum utilization of primary users (PUs), 
and determine the dynamic access by the secondary transmitters (STs) to the spectrum. 
Despite perennial efforts since its inception, 
the current DSM only works in a relatively simple RF environment with a small number of connected devices for a particular service~\cite{Advances_spectrum_sensing}. 
However, 
future massive connectivity will lead to an extremely complex RF environment, 
featured by various applications with different quality of service (QoS) requirements, 
a tremendous number of nodes, 
fast-changing traffic pattern dynamics, 
spectrum heterogeneity across the network topology, 
and increasing interference. 
These features significantly strain the current wireless systems, 
and hence call for the development of a revolutionary intelligent radio paradigm.

\begin{figure*}[htbp]
\centering
\includegraphics[width=0.65\textwidth]{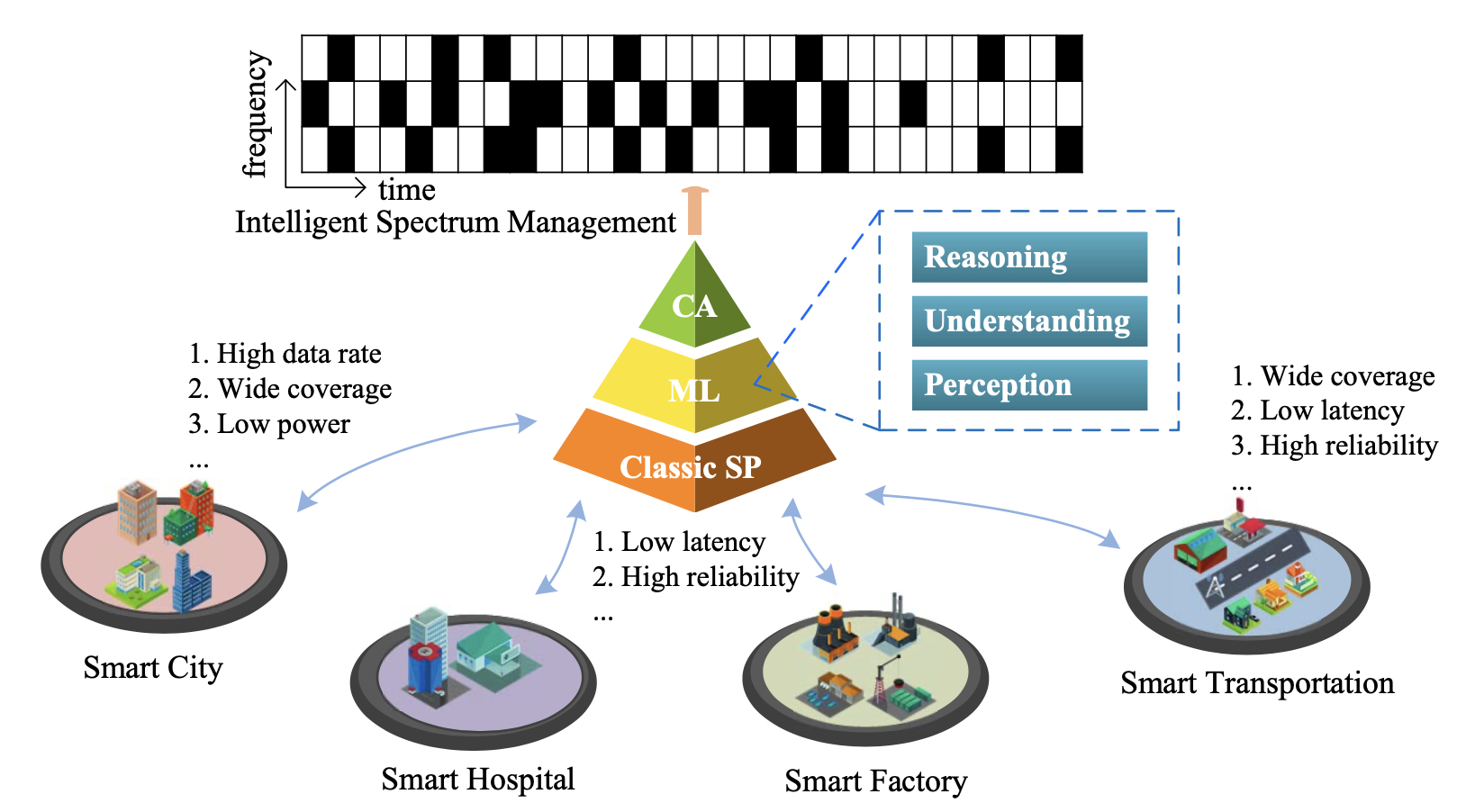}
\caption{A high-level schematic picture for intelligent radio evolution. The intelligence level increases from bottom to top of the pyramid. Numbered lists refer to user requirements. Above the pyramid, the black stripe refers to the occupied frequency while the white unoccupied. }\label{fig_8}
%\vspace{-4mm}
\end{figure*}

In~\cite{CR}, 
Haykin envisioned future intelligent radios as ``brain-empowered wireless devices". 
Considering the various underlying technologies and different levels of intelligence, 
%it is not surprising to see discrepancy in intelligent radio classification. 
%In this article, 
%we follow the line of intelligence level, 
%and 
in this article we propose to categorize spectrum intelligent radio into the following three streams, 
as illustrated by the intelligence pyramid in Fig.~\ref{fig_8}:
\begin{itemize}
    \item (S1) Human-oriented classical signal processing, 
    where sets of rules are created to represent knowledge in well-defined domains; 
    \item (S2) Machine learning (ML), 
    where statistical models are created for specific problem domains and trained from big data; and
    \item (S3) Contextual adaptation (CA), 
    where contextual explanatory models are developed to represent the real-world phenomena. 
\end{itemize}
In S1, 
the classical signal processing approach carries limited intelligence, which is well covered in the literature. 
As a leap forward, 
ML in S2 could be regarded as network deployments with partial intelligence.
It represents a new and significant trend, 
and starts to capture the current primary interest in the research community. Therefore, 
we focus on S2 in this article. 
Finally, 
S3 with CA represents a future-advanced stage of ML development with intelligence in its complete capacity. 
From another perspective, 
S3 achieves a machine-speed (i.e., real-time) decision on spectrum management, 
as an agile response to the large number of parameters involved in the complicated RF environments.  It is believed to provides a potentially complete solution to address the spectrum challenges in future massive network connectivity.

% Because of the burgeoning demand for spectrum, human approaches and timescales simply are not efficient enough. If we want to eek all of the utility out of the spectrum, we don't want to make decisions about how it s used on a timescale of hours or days. We want to make the decisions in seconds and milliseconds, asserted. We want to make machine-speed decisions on how we how we use the wireless spectrum.

In this article, 
we will introduce some of our recent applications and contributions of ML techniques in intelligent radio. 
This research area is at its early stage. 
However, 
our results and approaches  reveal promising future of applying ML techniques to achieve high levels of spectrum intelligence. 
Regarding this, 
we will present our vision for the research direction and expectation beyond the horizon. 
In general terms, 
ML enables machines to automatically learn from data and is the key technology behind self-driving cars and 
face recognition.
These successful applications motivate the embrace of the ML techniques, 
%the embrace of the rich body of existing knowledge in machine learning, 
leveraging the interdisciplinary synergy. 
By analogy to the recognition process of human brain, 
we divide the whole process of ML-based technology into three hierarchical levels: 
\textbf{perception}, \textbf{understanding}, and \textbf{reasoning}, 
illustrated in Fig.~\ref{fig_8}. 
\textbf{Perception} autonomously identifies  multiple  features of   signals, 
while \textbf{understanding} learns the intricate structure of the RF environment in a large-scale complex network, 
and establishes the ongoing RF activity map. 
At the highest level, \textbf{reasoning} builds upon perception and understanding, 
and reasons how to seamlessly access the shared spectrum resources with minimal interference to the PU. 
These three levels are relevant to different aspects of ML, 
and collectively enable the SU to ``see" the ongoing RF activities and understand the in-depth wireless landscape.  Specifically, the proposed perception method achieves fully blind multi-level spectrum sensing with an accuracy close to 1 when signal-to-noise ratio (SNR) is -12~dB. The understanding method can predict the primary users' coverage radius with an error of 2.8\%, and the reasoning method significantly outperforms the existing methods and approaches the optimal one in terms of idle channel selection accuracy.

\section{Stream 1: Human-Oriented Classical Signal Processing} 
%Some design/performance metrics that spectrum sensing and decision making aim for can be mentioned.% Add more examples. 
The majority of relevant research falls into the first stream based on human-oriented classical signal processing. 
Its typical feature is that a human being programs systems with explicit rules or logic (e.g., energy detection threshold) to implement \textit{handcrafted knowledge} in a very limited number of domains. 
In reality, 
classical signal processing is a complicated multi-task process. 
Spectrum sensing and decision making are the two most important tasks. 
%Sensing aims to minimize collisions with (or interference to) PUs, 
%while decision making explores possible opportunities and coordinates multiple SUs to share the opportunities considering fairness and efficiency in a distributed manner.

\subsection{Spectrum Sensing}
Spectrum sensing is a physical operation that measures the surrounding radio spectrum states based on various signal processing methods, 
including matched filtering, 
cyclostationary, 
energy detection for narrow-band scenario, 
and sub-Nyquist techniques for wideband case. 
However, 
these methods solely focus on a single parameter, 
the occupancy of a specific spectrum. 
In general, 
false alarm and detection probability are the most essential performance metrics. 
It is clear that a rich ensemble of RF activities and variations in the physical world are ignored. 
On the other hand, 
these methods usually assume a homogeneous spectrum state, 
where the propagation range of a PU is large enough to cover all SUs, 
and the whole network shares one common sparse spectrum~\cite{Huang}. 
This single-parameter sensing approach, 
combined with the unrealistic assumption on homogeneity, 
is incapable of handling complex RF environments of future massive connectivity.

\subsection{Decision Making}
Decision making consists of a series of logical operations, 
including sensing scheduling (when to perform sensing and which channels to sense), 
and channel access strategy (whether and which channels to access based on the spectrum sensing results). 
In addition, 
when only a small portion of the spectrum can be sensed at one time by low-cost battery-powered devices, 
spectrum sensing and channel access strategies can be treated jointly under a unified framework. 
Conventional studies use model-dependent approaches such as myopic (greedy) strategy, 
classical game theory, 
congestion control, 
matching theory, 
and graph coloring \cite{graphic} to obtain structured solutions. 
%For example, 
%the channel allocation problem can be translated into a graph coloring problem with vertice-user and color-channel pairs~\cite{graphic}.
Essentially, 
the model-dependent methods tackle a global constrained optimization problem by decomposition/decoupling, 
requiring the knowledge of the \textit{a priori} parameters in the network, 
such as user activation dynamics in certain spectrum. 

In practice, 
the complexity of spectrum environment often makes it impossible to gain enough knowledge in advance to develop the network evolution model, 
especially for the fast-changing RF dynamics, 
whose statistics is costly to obtain accurately due to its inherent complexity and practical limitations such as sensing time, 
hardware capability, and computational power. 
Furthermore, 
in the envisioned mMTC applications, 
decision making should cater for a decentralized self-organized multi-agent (multiple SUs) scenario, 
which is extremely challenging. 
All these limitations urge us to look for model-free decision-making~\cite{Model-Free}. 
A model-free method helps a decision-making entity interact with the environment, 
eliminating the need for a pre-defined network model. 
To achieve this, 
 techniques more advanced beyond handcrafted approach are required, 
resorting to the contemporary interdisciplinary techniques, 
such as ML-based approaches to directly interact with the unknown RF environment.

\section{Stream 2: Machine Learning}
ML automatically learns a statistical model of a specific problem using Big Data analytics. 
Supervised, unsupervised, and reinforcement learning (RL) are three commonly used ML algorithms. 
``Supervised/unsupervised" ML algorithms highlight whether the data samples are manually labeled or not, 
while reinforcement learning uses a trial and error methodology, 
which maps states to actions to maximize a numerical reward by a combined exploitation and exploration strategy. In general, supervised/unsupervised learning is efficient in understanding the PU's behavior, while reinforcement learning facilitates devices' access coordination. 
%In practice, it is difficult to label data due to the lack of prior knowledge on RF environments. 
%This makes unsupervised and reinforcement learning viable options. 
On this basis, 
we propose a new architecture for ML-based technology with three hierarchical levels. This enables an SU to autonomously  perceive, understand, and reason the unknown environments. 
%For specific applications, in this article perception targets the use-case of the PU with multiple power levels, the common signalling strategies in many standards. Understanding targets the use-case of a large-scale heterogeneous network, a feature of future complex networks. Finally, reasoning targets the use-case of spectrum access for correlated channels with unknown channel dynamics, a feature of the most realistic channels. It is worth noting that a trade-off always holds between the time taken for spectrum learning and system performance. However, learning time is an one-off operation, unless the radio environment changes significantly. 

%On a high level, perception is to identify the cross-layer multiple parameters that characterize the complex RF signals under a unified framework. On this basis, understanding is to learn and understand the intricate structure of the RF environment in a large-scale complex heterogenous network. Finally, reasoning is to achieve almost-instantaneous intelligent decisions on when, where, and how to optimally use shared spectrum resources among a large number of devices with different quality of service (QoS) requirements. 

\subsection{RF Landscape Perception}
The first level involves the autonomous multiple feature identification of signals in an unknown complicated RF environment. 
This allows an SU to observe network heterogeneity and dynamics from different perspectives. 
It extends a single feature identification in Stream 1 to a multiple one. 
 
\begin{figure*}[!tbp]
\centering
\includegraphics[width=0.65\textwidth]{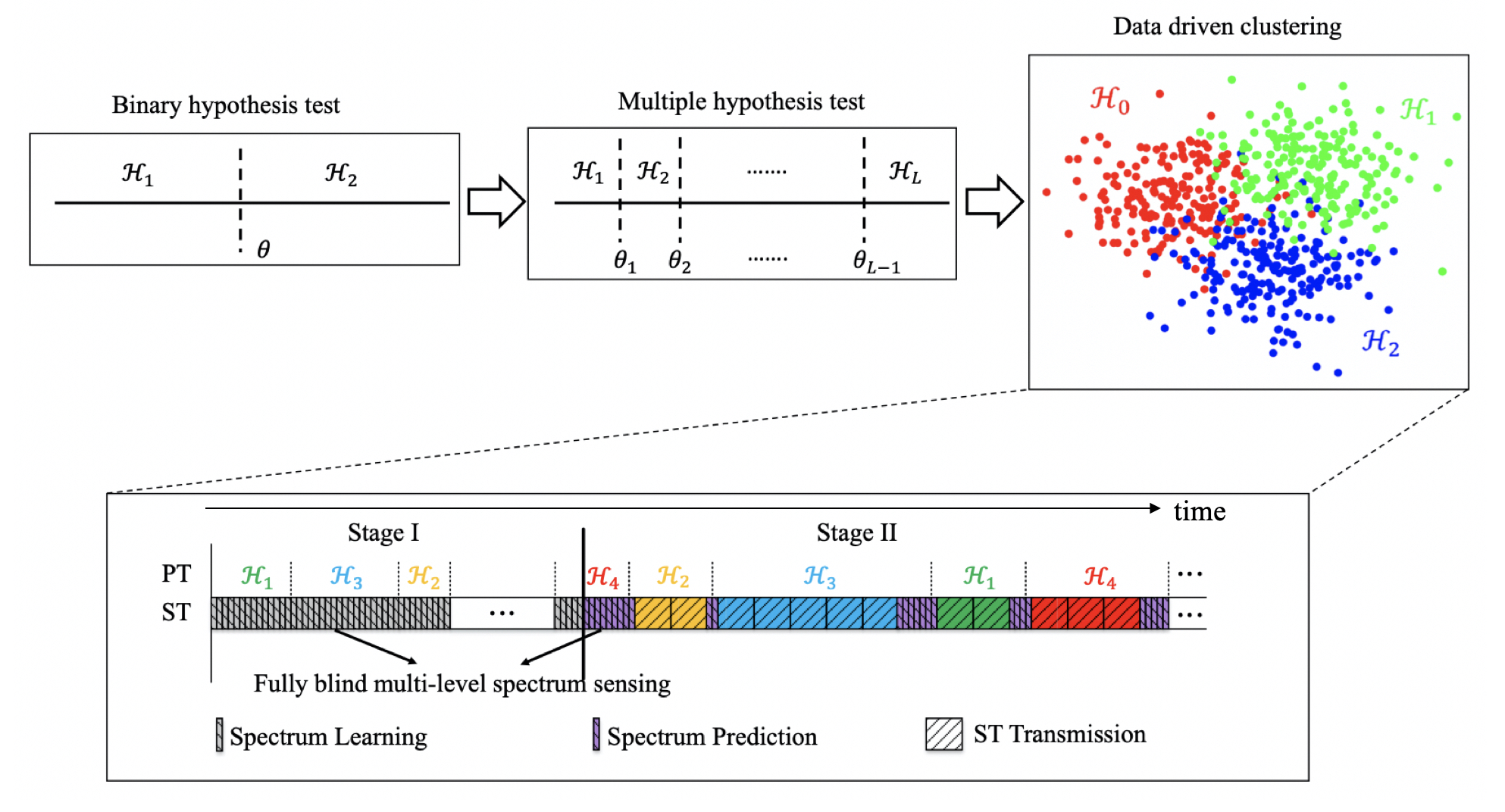} 
\caption{Multi-parameter cognition in RF landscape perception. The upper part  of the figure illustrates the evolution of methods for multi-level spectrum perception. Binary hypothesis test reports the primary transmitter (PT) as ON/OFF, while multiple hypothesis test identifies the exact PT power levels but with impractical requirement on prior information. Our proposed data-driven clustering method needs no such prior information, and automatically clusters the signals with the same PT transmit power level.    }\label{multilevel_simulation}
%\vspace{-4mm}
\end{figure*}

As discussed before, 
the spectrum sensing in Stream 1 mainly 
%involves two kinds of errors, and the sensing goal is to 
determines a detection threshold $\theta$ shown in Fig.~\ref{multilevel_simulation}, 
which is readily solvable. 
For example, 
given the target probability of false alarm and the noise power, 
$\theta$ can be simply determined by the Neyman-Pearson criterion. 
The work in~\cite{feifei} lifts this barrier, 
and considers different PU signal power levels as a new feature, 
potentially enabling the design of a more efficient spectrum access strategy. 
The goal of the \textit{multi-level} case is to jointly determine multiple thresholds $\{\theta_i\}_{i=1}^{L-1}$ to differentiate $L$ multiple power levels, 
which is far more complicated than the binary one. 
In essence, 
there are $L(L-1)$ possibilities of errors that are intertwined to exacerbate the complexity in threshold calculation. 
In~\cite{feifei}, 
closed-form thresholds were determined by assuming the same cost for all the errors. 
Note that a number of parameters (power mode) are required \textit{a priori}, 
including the number of power levels, 
the value of powers, 
the noise power, 
and the prior probability of hypotheses. 
In theory, 
this problem can be solved by labelling the collected samples from different PU power levels~\cite{feifei}. 
However, this is not very practical in reality. 

To remove the limitation of the existing work and achieve multi-level spectrum sensing with no or minimal prior information, 
we recently proposed a data-driven/ML based scheme~\cite{rui_level}. 
It is fully blind in the sense that the SU does not require any prior knowledge of the power mode. 
Specifically, 
the proposed spectrum sensing scheme spans across two stages as shown in Fig.~\ref{multilevel_simulation}. 
In Stage I (spectrum learning, a.k.a the training phase in ML), 
the SU collects a multitude of signals long enough to ensure
all the possible PU transmit power levels have been operated with a high probability. 
Then, 
a Gaussian mixture model (GMM) is used to capture the multi-level power characteristics of the signals. 
Finally, 
we introduce a Bayesian nonparametric method, 
referred to as conditionally conjugate Dirichlet process GMM (CCDPGMM), 
to automatically cluster the signals with the same PU transmit power and infer the model parameters (GMM parameters and PU power level duration distribution parameters). 
With the model parameters inferred in Stage I, 
the prediction part in Stage II can easily identify the current PU power level from the collected PU signal samples. 
In this way, 
Stage I together with the prediction part in Stage II achieve fully blind multi-level spectrum sensing. 
%
%
%
% %       
%\begin{figure}
%	\centering
%	\subfigure[Spectrum learning: from hypothesis test to blind data driven.]{
%		\includegraphics[width=0.4\textwidth]{Data_Drive}
%		\label{binary_mul_data}
%	}
%	\subfigure[The proposed two-stage multi-level spectrum sensing strategy. The sensing slots of the SU in both stages have the same time duration. The sensing slots in Stage I are used for learning, while that in Stage II are for prediction.]{
%		\includegraphics[width=0.5\textwidth]{Operation_Model}
%		\label{multilevel_simulation_right}
%	}
%	\caption{Multi-parameter cognition in RF landscape perception.}\label{multilevel_simulation}
%\end{figure}

%
%
%
%
%\begin{figure}
%	\centering
%	\subfigure[$N_s=10^4$.]{
%		\includegraphics[width=0.46\textwidth]{FF_0728_cluster_dB-eps-converted-to.pdf}
%		\label{Pc_left}
%	}
%	\subfigure[$\gamma_{st}=-12$ dB. ]{
%		\includegraphics[width=0.46\textwidth]{FF_0728_cluster_Ns-eps-converted-to.pdf}
%		\label{Pc_right}
%	}
%	\caption{The probability of correct PU power level prediction in the second stage ($P_c$) versus $\gamma_{st}$ and $N_s$.}\label{Pc}
%\end{figure}

\begin{figure}[!tbp]
\centering
\includegraphics[width=0.5\textwidth]{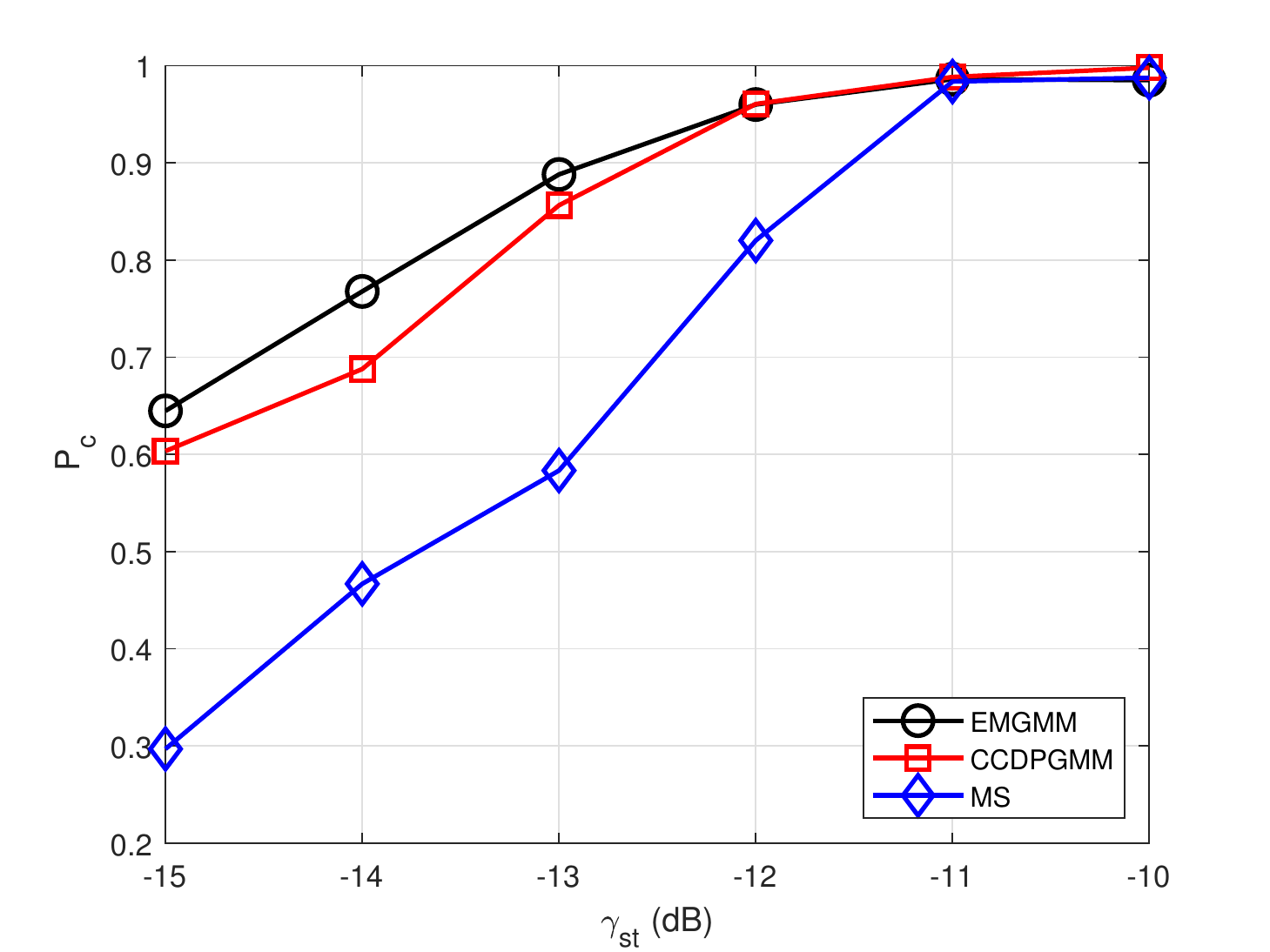} 
\caption{The probability of correct PU power level prediction in the second stage ($P_c$) versus $\gamma_{st}$. An additive white Gaussian noise channel is assumed for the purpose of illustration.}\label{Pc}
%\vspace{-4mm}
\end{figure}

The simulation results are presented in Fig.~\ref{Pc} to compare the accuracy of PU power level identification $P_c$ for the proposed CCDPGMM, 
expectation maximization GMM (EMGMM), 
and mean shift (MS) with respect to the average received SNR $\gamma_{st}$. 
The EMGMM is a parametric clustering method that requires the prior knowledge of $L$, 
while the CCDPGMM and MS belong to a nonparametric class without the need for such prior knowledge. 
We set the power level $L = 4$ and the probability of each hypothesis ${\rm P_r} (H_l)$ = 0.25. 
It is assumed that the noise variance is 1, 
the PU transmit powers $P_1 : P_2 :P_3 =1:2:3$, 
and $P_4 =0$. 
It is shown that the proposed CCDPGMM significantly outperforms MS (for example, a 2.2 dB gain at $P_c=0.6$), 
and is only slightly inferior to EMGMM.  At $\gamma_{st}=-12$~dB, CCDPGMM achieves $P_c=0.96$.  Note that the accurate estimation of the power levels is important, or else it will degrade the performance in Stage II. 

%It is shown in Fig. \ref{multilevel_simulation_left} that $P_c$ increases with $\gamma_{st}$ for all the three methods. This is because the gap between the adjacent transmit powers increases with $\gamma_{st}$, rendering them more distinguishable by machine learning. Similarly, $P_c$ increases with $N_s$ in Fig. \ref{multilevel_simulation_right}, because a larger $N_s$ results in a smaller variance of each Gaussian distribution in the mixture model. Furthermore, without the prior knowledge of $L$, CCDPGMM outperforms MS, significantly so for small $\gamma_{st}$ and large $N_s$. The accurate PU power level identification lays a solid foundation for smart spectrum access design. 

Clearly, 
in order to achieve a full RF cognition, 
future networks demand automated extraction of far more features with no or minimal prior information. 
Preferably, 
the physical layer information (spectrum occupancy, transmit power level, modulation, constellation, and channel coding) and upper layer features (application types, network topology, and communication protocols) should be mined under a unified framework. 
In this line, 
a deep convolutional neural network (CNN), 
built on the layers of convolving trainable filters, 
can perform nonlinear approximations and emerge as a tool of the future to automate the extraction of a multitude of features. 
This represents a new trend for RF landscape perception.

\subsection{RF Environment Understanding}
The second level learns the intricate structure of the RF environment in a large-scale complex network, 
and establishes the ongoing RF activity map. 
This may include understanding large and small-scale wireless channel propagation conditions and identifying non-collaborative wireless systems. 
Network complexity is mainly embodied in heterogeneity and dynamics. 
For example, 
a future network could include multiple PUs across a large area, 
and the spectrum states could vary significantly from place to place within the network. 
This heterogeneous property stymies the accurate spectrum sensing and availability prediction for a given location. 
Considering the potentially large number of spectrum states, 
one can deploy many static SUs at different geographic locations to carry out spectrum sensing simultaneously, 
and then collate all the sensing results to identify all the different spectrum states. 
However, 
this is highly undesirable due to the high deployment cost.  

\begin{figure*}[htbp]
\centering
\includegraphics[width=0.65\textwidth]{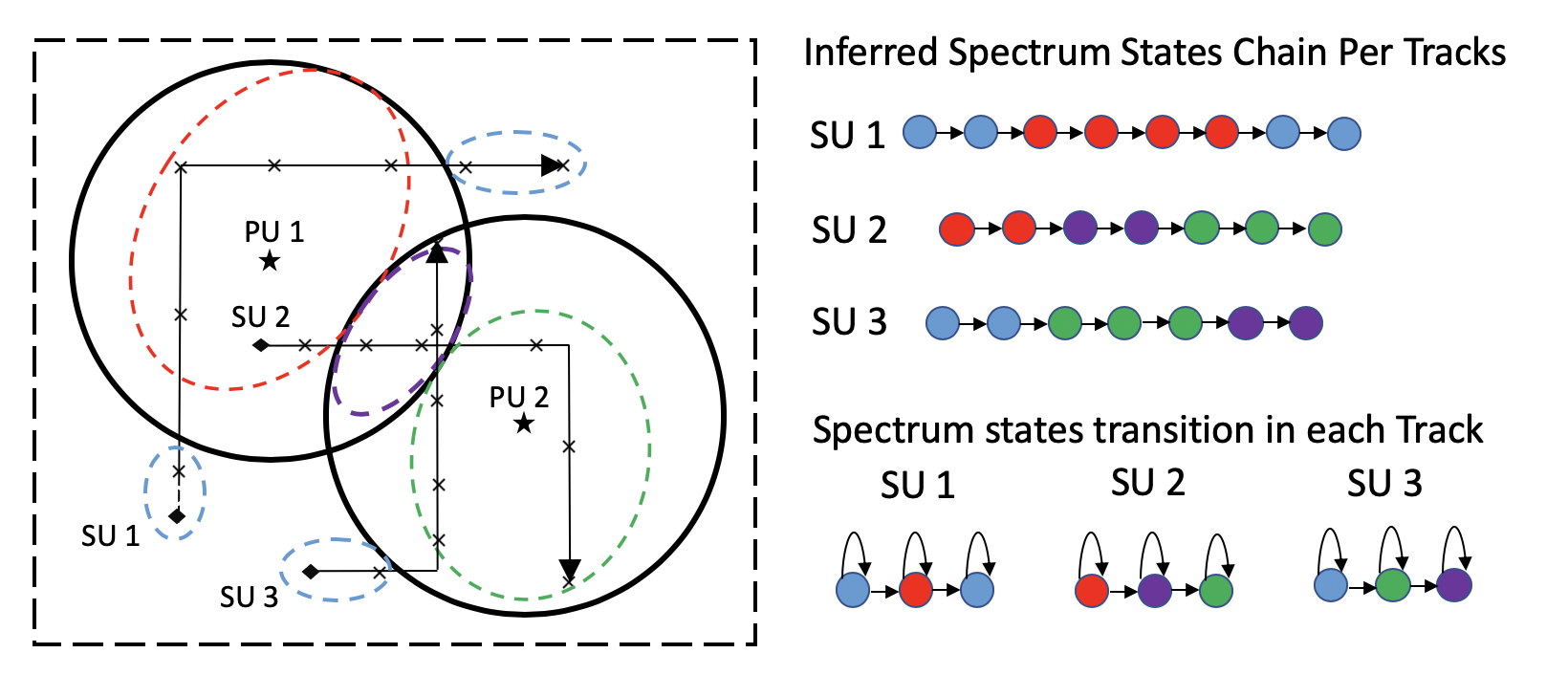}
\caption{The basic concept illustration of the proposed mobile collaborative spectrum understanding. Two black stars represent two PUs, and two black  circles the transmission ranges. Black straight lines represent the tracks of SUs. The tracks are divided into four different sections indicated by different colors, each color representing a unique inferred spectrum state. Black crosses mark the random sensing locations, and in the inferred spectrum state chain, each point corresponds to one of these locations. Spectrum state transition is obtained by combining the consecutive same color points.} \label{fig_illustration}
\end{figure*}

To understand the RF environment in a more cost-efficient way, 
we exploit the mobility nature inherent to most wireless devices to explore the spectrum footprint across a network~\cite{Yizhen}. 
Specifically, 
a small number of SUs simultaneously sample the RF channels while moving along, 
and each SU sends its sensing results to its nearby cluster head (CH). 
These CHs are interconnected and exchange information with each other, 
and thus the global spectrum states are derived cooperatively. 
This basic collaborative concept is illustrated in Fig.~\ref{fig_illustration}.  
Mathematically, 
we proposed a beta process sticky hidden Markov model (BP-SHMM) to represent and capture spatial-temporal correlation among RF samples. 
Then, 
Bayesian inference is adopted to classify sensing samples into different classes in an unsupervised manner. 
In essence, 
all the spectrum samples in each class have the same spectrum state. 
Then the locations of PUs and their transmission ranges can be inferred from the classification results. In specific, our calculation indicates that the coverage radius prediction error is 2.8\%.
Consequently, 
a newly joined SU could have almost instantaneous access to the available frequency channels without sensing. 
The accuracy is much higher than individual SU sensing, 
and it also brings an important benefit of low latency. 
\begin{figure}[htbp]
\centering
\includegraphics[width=0.4\textwidth]{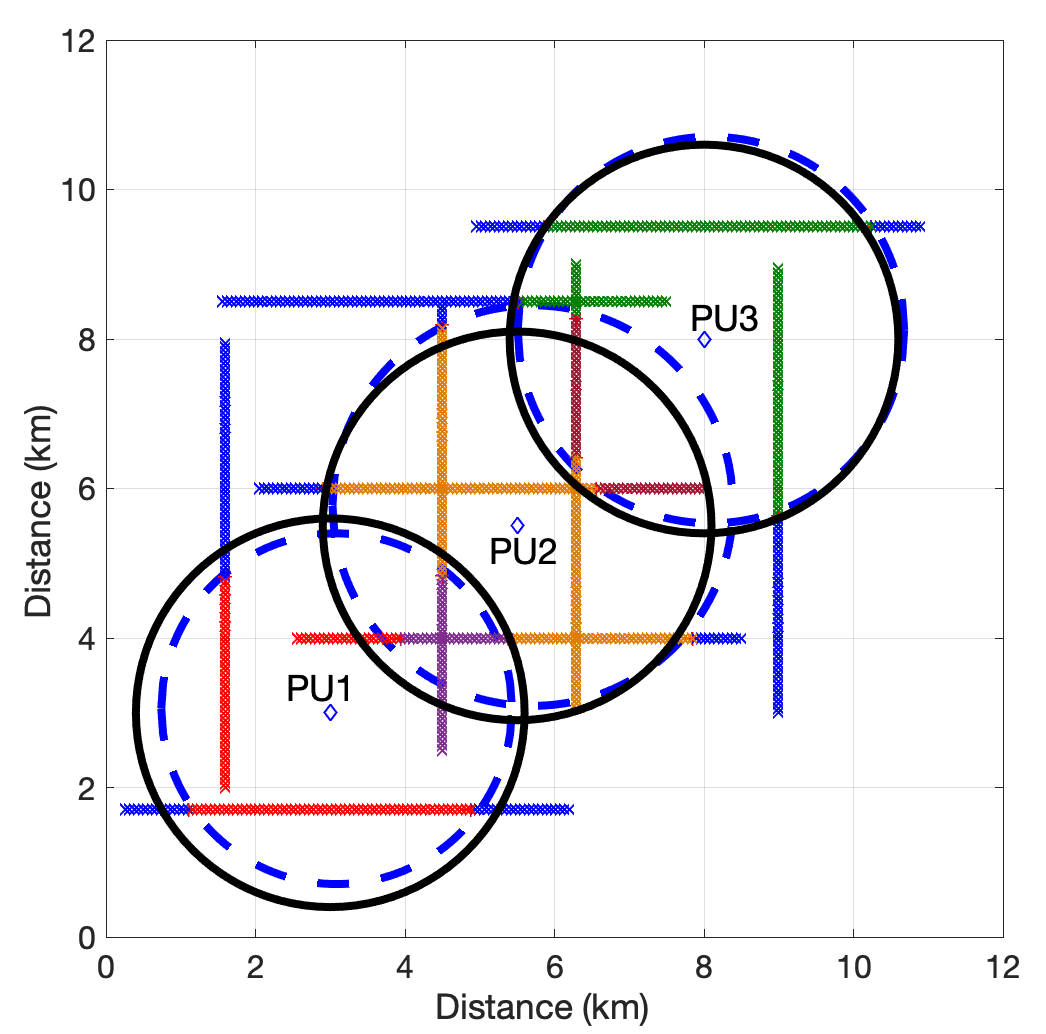} 
\caption{An example of classification and prediction results in a network ($12 \times 12$ km) with three active PUs and nine mobile SUs each with a straight track. Black circles indicate the PUs' actual transmission ranges (radius of 2.2 km), while blue-dashed circles the predicted coverage. Different colors on the track refer to different inferred spectrum states (classification).}\label{fig_fittingresults}
%\vspace{-4mm}
\end{figure}
The superiority of our proposed method is illustrated by an example in Fig.~\ref{fig_fittingresults}. 
It is clearly shown that all the six spectrum states (each represented by a different color) were identified successfully and the coverage areas predicted,
both with high accuracy. 
%We define the relative radius error as $\frac{r_{p}-r_{r}}{r_{r}}$, where $r_{p}$ is the predicted radius and $r_{r}$ is the real radius. We find that the average value across the three PUs is $2.8\%$, an indication of a high inference accuracy.

The proposed method mainly focuses on the spectrum heterogeneity. 
Handling the envisioned scenario with fast-changing dynamics and interference is still an open problem. 
A potential way is to design two coupled SHMMs. 
The first one models the latent statistical correlation within each mobile SU's sequential sensing samples in terms of the physical layer parameters, 
while the second one targets the application layer parameters interpreting the dynamics from an application perspective and identifying interference. 
This may allow us to gauge the mutual influence of the physical layer and application layer parameters simultaneously.

\subsection{Reasoning for Instantaneous Spectrum Access }
At the third level, 
 SU can perceive its surrounding spectrum environment and make use of its understanding of environmental features to reason how to seamlessly access the spectrum resources shared among devices, 
minimizing the interference to the PU. 
Essentially, 
multi-user dynamic spectrum access is a highly challenging distributed decision-making problem~\cite{Decentralized}, 
as the network utility should be maximized without any online coordination or inter-user information exchange. 
As explained before,
when facing a complex dynamic network, 
model-dependent decision-marking approaches are no longer effective. 
In contrast,
model-free learning enables the SU to interact with the environment directly and then adjust its behavior by reinforcement learning, 
the key to realizing intelligent spectrum. 

\begin{figure}[!tbp]
\centering
\includegraphics[width=0.5\textwidth]{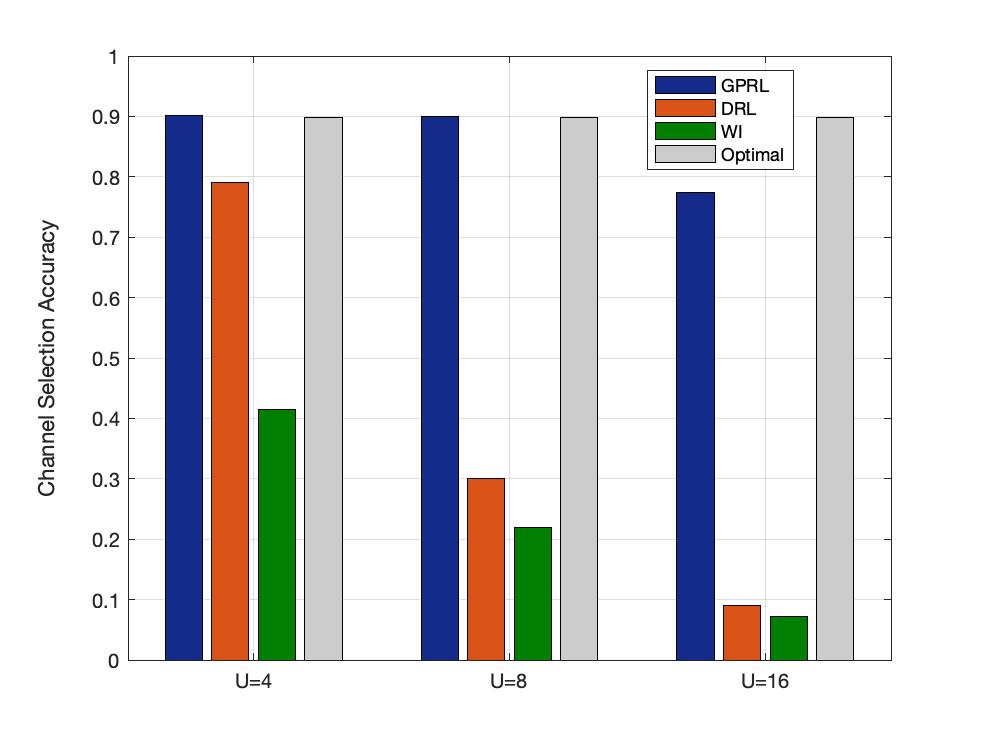} 
\caption{Channel selection accuracy comparison of different  methods for varying numbers of subsets. Sixteen channels are randomly divided into $U$ subsets, each containing $16/U$ channels which share  the same idle/occupied  state and state transition pattern (correlated channels). }\label{Figure-DRL}
%\vspace{-4mm}
\end{figure}

In practice, 
a battery-powered SU can only perceive a small portion of the available channels at an instance, 
and needs to select a possible idle channel(s). 
As such, 
dynamic multi-channel access problem can be modeled as a partially observable Markov decision process (POMDP)~\cite{Decentralized}, 
which is the generalization of an MDP. 
When the network dynamics (the state transition probability of the channel) is unknown and the channels are correlated, 
the POMDP becomes very challenging due to high-dimensional and large state-space issues. 
The popular reinforcement learning approaches, 
such as Q-learning, 
are intractable to solve it. 
To address this issue, 
the work in~\cite{DRL} uses a deep neural network (DNN) approximate Q-values, 
replacing the conventional look-up Q table. 
However, 
DRL usually requires a large number of training samples to tap the strong mapping capability of DNNs~\cite{Silver2016}.
However, 
only a single sensing sample for one channel can be collected in each time slot,
and it results in a low learning rate and channel selection accuracy. 

We propose a novel model-free Gaussian process reinforcement learning (GPRL) based solution. 
As a Q-function approximator in RL, 
the GP is embedded to measure  data correlation using a kernel function and update the posterior Q-value distribution via the Bayesian theory. 
On this basis, 
the past learning experience can be utilized in a more efficient way compared to the DNN-based Q-function approximators. 
This accelerates learning and eliminates the need for a large number of training samples. 
Fig.~\ref{Figure-DRL} compares the channel selection accuracy of the proposed GPRL, DRL, Whittle index (WI), 
and the optimal (ideal) method for a total of 16 channels. 
The results of DRL and GPRL are obtained via an online learning process with learning and testing phases of 120 and 30 time spans, 
while WI and the optimal method are directly applied in the testing phase. It is clearly shown that GPRL outperforms DRL significantly for increased $U$, and achieves an accuracy close to the optimal one.

However, 
GPRL and DRL only suit a single-user scenario. 
The multi-user setting is significantly different in environment dynamics, 
network utility, 
and algorithm design, 
which is much more challenging to solve than the single-user case. 
Due to interactions among users, 
it is highly desirable to develop a model-free distributed multi-user method without coordination or message exchange among users. 

%will determine the spectrum access by SUs.  After the training phase in Task 3.1, the SUs will only need to update their DRL weights by communicating with the central unit. In real time, at each time slot, each user maps its current state to spectrum access actions (when, where, and how to select next channel) based on a trained DRL to maximise the objective function. The DRL design will enable each SU to learn a good policy in a fully distributed manner, while dealing with large state space without online coordination or information exchanges between SUs. Training the DRL with different objective functions might lead to significantly different operating points of the system. I will analyse the system dynamics to establish the design principles for the implementation of the proposed algorithm. Under both cooperative and non-cooperative network utilities, I will develop new method to optimise the channel access based on Bayesian perspectives [29-30] to increase the channel throughput by further reducing the number of idle time slots and collisions. 

\section{Stream 3: Contextual Adaptation}

Motivated by the next generation of artificial intelligence\cite{holzinger2018machine, 8253597}, the third stream of spectrum intelligent radio has been envisioned to feature contextual adaptation, 
and meet the need for future massive connectivity with its full intelligence. 
It is paramount that the radio systems contextualize the acquired knowledge, 
and transfer the relevant information from the previous scenarios to the current one. 
Essentially, 
contextual explanatory models are developed to autonomously identify the key elements and capture corresponding parameters relevant to the explanation and representation of the real-world phenomena. 
In the context of spectrum intelligent radio, 
an SU autonomously reasons and learns from the sensed radio environment to make a decision on the communications mode most appropriate to the prevailing conditions, 
taking into account its performance targets. 
It is able to evaluate a variety of criteria, 
including delay and power constraints as well as sensing accuracy and security requirements, 
and selects the most practical one suiting the specific radio environment. Furthermore, it self-reconfigures its hardware to execute the selected communication mode.

Mathematically, 
contextual adaptation relies on a contextual model to represent the real-world settings, 
leveraging the interpretable machine-learning technique. 
That is, 
this model does not rigidly or blindly follow preset rules for decision making. 
Instead, 
the intelligent radio proactively discovers the logical rules and defines the models to shape the decision-making process. 
Consequently, 
it comprehends its actions and understands the rationale behind its decisions. 
For example, 
the intelligent radio does not only identify the best channel for SU access, 
but also establishes and understands the inherent relationship between its decision and DNN parameters. 
With a properly designed human-computer interface, 
the decision process can manifest itself to a human user. 
This enables the radio to truly ``see", perceive, 
and understand the ongoing RF activities,
as well as make transparent and trustworthy decisions. 
Moreover, interpretable ML can predict how it will behave in the future, 
removing the technical hurdles faced by the current ML algorithms, 
where a new scenario that can severely deviate from the training model renders radio malfunction.  
An intelligent radio with contextual understanding will entail far fewer samples to be trained, utilizing contextual models. 
In essence, 
such a system over time will learn about how a contextual model should be structured.
Once established, 
this model will be the basis for a radio system to reason the RF landscape and ongoing activities, 
as well as make decisions on RF access.
For massive device access necessary for the future dense IoT network featured in Industrial 4.0, 
contextual adaptation-enabled devices will autonomously sense a complex local RF landscape, 
determining how to avoid interference, 
and exploiting opportunities to intelligently and efficiently access the available spectrum. 
From another perspective, 
the current ML approaches, 
DNN in particular, 
work in a ``black-box" style and the decisions of the ML models are not intuitive and usually inexplicable to human users. 
Consequently, 
trust in the system can not be firmly established and proved. 
This is a genuine concern in the context of Industrial 4.0, 
where all the control signaling must be conducted in a clearly understandable fashion. 
Therefore, 
the ``explicable" ML calls for theoretic breakthroughs, 
which continues to pose a grand challenge for the entire scientific community.

\section{Development Roadmap}

Compelled by the overwhelming impact of spectrum scarcity on the national digital economy, 
many government regulators such as FCC in the U.S., Ofcom in the U.K., and ACMA in Australia, 
have started to rethink radio frequency management to create a connected digital society.  
Industry and standardization initiatives, 
under the auspices of major regulatory agencies, 
have been mobilized to bring such management concept into standardization,
including the early standard version of IEEE 802.22 and new standards such as IEEE 802.11af and ECMA 392. 
Noteworthy, the above standards have been designed for the TV white space, 
where a device can obtain an available channel list from TV white space database. In the future, it is highly desirable to see the standardization process on other frequency bands. 

The Defense Advanced Research Projects Agency (DARPA) plays a leading role in developing ML-based spectrum management.
The recently announced Spectrum Collaboration Challenge (SC2) aims to develop and integrate advanced ML capabilities into the current radio applications to achieve a collective optimization of the wireless spectrum strategies. 
The ultimate goal targets the dynamic and instantaneous spectrum sharing at machine timescales. 
In fact, 
in an earlier demonstration, 
the ML-enabled radios were seen to have outperformed human spectrum managers by 50\%. Recently, DARPA has also launched the radio frequency machine learning systems (RFMLS) program to address the performance limitations of conventionally designed RF systems such as radar, signal intelligence, electronic warfare, and communications. FCC has recently ushered in a test program to allow civilian users (LTE providers and general users) to share spectrum with the U.S. Navy on a bandwidth of 150 MHz at 3.5\,GHz, and ML has become a viable solution \cite{FCC3.5}.  
It is highly encouraging to see regulatory bodies are realizing the importance of dynamic spectrum management, 
and embracing the concept of spectrum intelligent radio. 
It is expected that only through the promulgation of governmental law and policy can such a concept be widely accepted and adopted by the broad industry. 

\section{Conclusion}
This article has discussed the background, progress, challenges, 
and future research trends of intelligent radio technologies. 
We elaborated on the three streams of intelligence development, 
and focused on the second stream ML, 
which has recently attracted great interest in the research community. 
We presented new results of our recent work on spectrum perception, understanding, and reasoning.  The government's regulatory impact was also discussed. 
We hope that this article stimulates more interest in this promising research area, 
and encourages more efforts are made toward the fully intelligent spectrum resource utilization. 
%We expect that, with a coherent evolution from the aspects of regulatory reform, 
%technical breakthrough, 
%and business model innovation,
%a fully intelligent spectrum resource management will become a reality in the future.

\bibliographystyle{IEEEtran}
\bibliography{Spectrum_AI_magazine_V7}

\begin{IEEEbiographynophoto}{Peng Cheng}
received the Ph.D. degree from Shanghai Jiao Tong University, China, in 2013. From 2014 to 2017, he was a postdoctoral research scientist at CSIRO, Sydney, Australia, and he is currently with the University of Sydney, Australia. His research interests include machine learning, wireless AI, network MIMO, OFDM and compressive sensing theory.
\end{IEEEbiographynophoto}
\begin{IEEEbiographynophoto}{Zhuo Chen}
received the B.S. degree  from Shanghai Jiao Tong University, China, and the Ph.D. degree from the University of Sydney,  Australia.
He was a Senior Research Scientist with CSIRO, Sydney, Australia. His research interests include wireless sensor networks, wireless AI, and signal processing.
\end{IEEEbiographynophoto}
\begin{IEEEbiographynophoto}{Ming Ding}
is a Senior Research Scientist at Data61, CSIRO, Australia. He has authored over 80 papers in IEEE journals and conferences,  and about 20 3GPP standardization contributions. He holds 14 US patents and co-invented another 100+ patents on 4G/5G technologies in CN, JP, EU, etc. Currently, he is an editor of IEEE Transactions on Wireless Communications.
\end{IEEEbiographynophoto}
\begin{IEEEbiographynophoto}{Yonghui Li}
 received his Ph.D. degree in 2002 from Beijing University of Aeronautics and Astronautics. Since 2003 he has been with the Centre of Excellence in Telecommunications, the University of Sydney, Australia, where he is now a professor.  His current research interests are in the area of wireless communications, with a particular focus on MIMO, millimeter wave communications, machine to machine communications, coding techniques and cooperative communications.
\end{IEEEbiographynophoto}
\begin{IEEEbiographynophoto}{Branka Vucetic}
is an ARC Laureate Fellow and professor of telecommunications, Director of the Centre of Excellence in Telecommunications at the University of Sydney. During her career, she has held research and academic positions in Yugoslavia, Australia, U.K. and China. Her research interests include coding, communication theory and signal processing and their applications in wireless networks and industrial Internet of Things. She is a Fellow of the Australian Academy of Technological Sciences and Engineering and a Fellow of the IEEE.
\end{IEEEbiographynophoto}

\begin{IEEEbiographynophoto}{Dusit Niyato}
received the B.Eng. degree from the King Mongkut’s Institute of Technology Ladkrabang, Thailand, in 1999, and the Ph.D. degree from the University of Manitoba, Canada, in 2008. He is currently a Professor with the School of Computer Science and Engineering, Nanyang Technological University, Singapore. His research interests are in the areas of energy harvesting for wireless communication, the Internet of Things, and sensor networks.
\end{IEEEbiographynophoto}

\end{document}